# Amplifier for scanning tunneling microscopy at MHz frequencies

K.M. Bastiaans [1], T. Benschop[1], D. Chatzopoulos[1], D.H. Cho[1], Q. Dong[2], Y. Jin[2], M.P. Allan[1]

1 Leiden Institute of Physics, Leiden University, Niels Bohrweg 2, 2333 CA Leiden, The Netherlands
2 Centre de Nanosciences et de Nanotechnologies, CNRS, Univ. Paris-Sud, Univ. Paris-Saclay, C2N – Marcoussis, 91460 Marcoussis, France

*Conventional scanning tunneling microscopy (STM) is limited to a bandwidth of circa 1kHz around DC. Here, we develop, build and test a novel amplifier circuit capable of measuring the tunneling current in the MHz regime while simultaneously performing conventional STM measurements. This is achieved with an amplifier circuit including a LC tank with a quality factor exceeding 600 and a home-built, low-noise high electron mobility transistor (HEMT). The amplifier circuit functions while simultaneously scanning with atomic resolution in the tunneling regime, i.e. at junction resistances in the range of giga-ohms, and down towards point contact spectroscopy. To enable high signal-to-noise and meet all technical requirements for the inclusion in a commercial low temperature, ultra-high vacuum STM, we use superconducting cross-wound inductors and choose materials and circuit elements with low heat load. We demonstrate the high performance of the amplifier by spatially mapping the Poissonian noise of tunneling electrons on an atomically clean Au(111) surface. We also show differential conductance spectroscopy measurements at 3MHz, demonstrating superior performance over conventional spectroscopy techniques. Further, our technology could be used to perform impedance matched spin resonance and distinguish Majorana modes from more conventional edge states.*

**I. Introduction & motivation**

Possible applications of scanning tunneling microscopy experiments in the MHz regime include high-frequency differential conductance measurements, scanning spin resonance experiments, and noise spectroscopy on the atomic scale. Conventionally, this is prevented in STM by the combination of a GOhm resistance of the tunnel junction and a capacitor from the cabling which form a low pass filter in the kHz regime. In this paper, we build a matching circuit including superconducting inductors and a home-built HEMT that allows us to measure STM currents at MHz frequencies while remaining in tunneling and with atomic resolution. We demonstrate the amplifier's superior performance for both scanning noise spectroscopy and MHz differential conductance measurements.

We start with an introduction to noise spectroscopy. Measurements of electronic noise can yield information in mesoscopic systems that is not present in their time-averaged transport characteristics, including fractional charges in the quantum hall regime[1,2], the doubling of charge in Andreev processes[3], Coulomb interactions in quantum dots[4–7] and the vanishing of noise in break junctions at the quantum conductance[8]. Generally, the quantity of interest is the Fano factor *F* which measures the deviation of the noise from the Poissonian noise of independent tunneling events of electrons, $S_P$ = 2*eI*, with *e* the electron charge and *I* the current[9,10]. The Fano factor is then defined as the ratio between measured (*S*) and Poissonian ($S_P$) noise, *F* = *S* / $S_P$. For an uncorrelated electronic liquid, one expects *F* = 1; but one can imagine systems where the charge of the carriers is not equal to the electron charge ($q \neq e$) or where the electron flow is strongly correlated. In these cases, the Fano factor will not be equal to unity, i.e. the current noise will be smaller or larger than the Poissonian value even though the time-averaged value of the current will not be influenced. Resolving the noise with atomic precision might provide us with new information in systems with strong electronic correlations or charge aggregations that are not present in the mean current. This is our main motivation to combine noise measurement with scanning probe microscopy.

As for the application of MHz differential conductance measurements, we use a lock-in amplifier as it is done conventionally, but with 3MHz instead of the more common 400Hz – 1kHz. The clear advantage is that in this way, one can perform the spectroscopy measurement in a frequency window where 1/*f* noise is much lower. In addition to this, we can clearly separate the high and low frequency signal, thus it is, for example, easier to measure in feedback.

Bringing noise measurements to STM in the tunneling regime ($R_{\text{junction}}$ >~ 1GOhm) comes with unique challenges, which prevented any atomic resolution noise measurement in the tunneling regime thus far. The high impedance of the tunnel junction, formed by the few angstroms vacuum gap between the STM tip and sample, is the critical obstacle. Together with the capacitance of the interfacing coax cable, the junction acts as a low pass filter only allowing transmission of signals in the small frequency range[11]. Moreover, conventional amplifiers used in STM also have a limited bandwidth due to a large feedback resistor and unavoidable parasitic capacitances[12]. This conventional STM circuitry limits the bandwidth to detect the tunneling current from DC to around 1kHz (fig. 1a). Possible solutions to this are bootstrapping the amplifier[13–16], or impedance matching[11]. These enabled noise measurements in MOhm tunnel junctions[11,13,17,18], but a GOhm impedance as it is present in many STM experiments still leads to prohibitive losses in the matching circuit.

Here, we report on a new amplifier circuit that allows us to overcome these challenges. The requirements for our amplifier were: (i) the amplifier should not interfere with traditional STM measurements, (ii) it should work in the GOhm regime, (iii), it has to be possible to easily implement the amplifier in a commercial STM, (iv) it has to be compatible with UHV, implying low outgassing so that the system can be baked and ultra-high- (cryogenic) vacuum can be achieved.

Our key figures of merit are: (i) the low noise of the circuit, (ii) the most efficient separation of high and low frequency signals, and (iii) the highest possible Q factor of the resonator for highest amplification at 3MHz.

This article is structured as follows. Noise in a STM junction is discussed in section II. A block diagram of the newly developed system for noise-spectroscopy measurements in STM is presented in section III, followed by a discussion on the requirements for implementing such techniques in STM. Section IV describes the realization of this new amplifier. A demonstration measurement on an Au(111) surface is presented in section V. Differential conductance measurements with MHz voltage modulation are discussed in section VI.

**II. Noise sources in STM**

We start by considering the types of unwanted noise present in a STM setup: mechanical noise, thermal (Johnson) noise, amplifier noise, and flicker (1/f) noise.

They have distinguishable frequency dependences, as shown in figure 1a. First, flicker noise or 1/f noise, which is present in almost all electronic devices. The power spectral density of this low-frequency phenomenon is inversely proportional to the frequency and is related to slow resistance fluctuations modulated by temperature variations. Second, noise induced by mechanical vibrations transferred to the junction, where this mechanical noise is converted to current noise. Both noise sources are usually present in the range from DC to a few kHz, indicated by the blue shaded area in figure 1a. This emphasizes that the low frequency regime should be avoided and illustrates the disadvantages of the conventional STM bandwidth.

At higher frequencies, the current fluctuations are dominated by thermal noise and shot noise, both of which are informative about the sample. In principle, both phenomena are frequency independent (white noise), and thus are also present at lower frequencies, where the total noise power is dominated by the other contributions. Thermal (also called Johnson) noise is the thermodynamic electronic noise in any conductor with a finite resistance $R$; its power spectral density is constant throughout the frequency spectrum, $S = 4k_B T$, where $k_B$ is Boltzmann's constant and is temperature. Since thermal noise in a conductor is proportional to temperature, it can be lowered by reducing the temperature. It can be distinguished from shot noise at zero current, where the latter vanishes.

Our goal is thus to increase the bandwidth and move it to higher frequencies, all whilst retaining the conventional capabilities and staying in the tunneling regime.

### III. Amplifier and circuit

### III.a General idea

To achieve the requirements and goals of section I while avoiding the unwanted noise sources described in section II, we develop a resonance circuit based amplifier including a resonator-based bias-tee.

We follow the principle of amplifier circuits built for noise spectroscopy measurements in mesoscopic systems[19–22] but we modify it to work for high junction resistances in the GOhm regime and to be compatible with STM. Figure 1b shows a block diagram of the amplifier circuit combined with STM. First, a bias-tee (green) separates the low- and high-frequency signals coming from the STM junction. The low-frequency part is needed for the STM feedback loop, where the current is converted to a voltage by a transimpedance amplifier at room temperature. To separate the high frequency, one could use a bias-tee consisting of an inductor in one arm and a capacitor in the other one. However, as we still need a kHz bandwidth in the low frequency branch and as we want to minimize losses of the high frequency signal, we use a resonator based bias-tee.

The high-frequency part of the signal is then passed through the parallel RLC circuit (tank, indicated in purple fig. 1b), which converts current to voltage at the resonance frequency of the tank circuit $f_0 = \left(2\pi\sqrt{LC}\right)^{-1}$. The voltage over the tank circuit is detected by the gate of a high electron mobility transistor (HEMT, indicated in yellow fig, 1b) with very low input referred voltage[23,24] and current noise, operating at the base temperature ($T \sim 3.5$ K) of the STM. Through the transimpedance of the HEMT, the voltage fluctuations at its gate are converted into current fluctuations. These are measured over a 50 Ω resistor to finalize the impedance transformation. Note that while the voltage/current gain of the amplifier is of order unity, the gain of power is considerable. A 50 Ω coaxial line connects the amplifier circuit to a commercial 40dB current amplifier at room temperature. Finally, the signal line is terminated by the 50 Ω input impedance of the spectrum analyzer.

**III.b Circuit elements and printed circuit board design**

The heart of the circuit is built on a ceramic printed circuit board (Rogers Corp TMM10i, selected for the very low outgassing properties) as depicted in figure 2 and described below. Figure 2a shows the circuit schematics of the amplifier. The board is located close to the STM head, at the base temperature of the liquid He 4 cryostat (Unisoku USM1500). Figure 2b shows a photograph of the board mounted on the STM. The tank circuit is covered with a superconducting Niobium shield (inset shows the tank circuit beneath).

The input of the amplifier is connected to the STM tip via a coaxial cable (silver plated Cu mini-coax CW2040-3650F) with a total capacitance between inner and outer conductor of $C_w$=30pF. The bias-tee (indicated by green shading) and tank (purple shading) combination is formed by two home-built superconducting Niobium inductors $L_1$= $L_2$=66μH coupled by capacitors $C_c$=100pF (Murata GRM 0805-size surface mount). The low-frequency transmission of the bias-tee is shown in figure 3a, measured at low temperature. The flat transfer function in the frequency range of the Femto IV amplifier (1kHz, blue shaded are in figure 3a) ensures that this amplification scheme can be used for the STM feedback system.

The resonance circuit is formed by the self-resonance of the superconducting Nb inductors in combination with the coaxial cable $C_w$, providing a resonance frequency of 3.009MHz. Parallel self-capacitances of the Nb inductors are also shown in figure 2a, $C_1$=15pF and $C_2$=15pF. The Niobium inductors are made by cross-winding annealed Nb wire of 100 μm in diameter around a customized ceramic (macor) core. We choose superconducting Nb inductors to enhance the quality factor of the resonator, increasing current-to-voltage amplification at resonance. At 4K, the Nb inductors show a high quality factor of Q=600, 50 times larger than similarly made Cu inductors (Q=12), see figure 3b. The Nb inductors are covered by a Nb shield to minimize Eddy current damping, ensuring the highest possible quality factor.

The high-impedance part of the amplification scheme (tank circuit coupled to STM junction) is matched to the 50 Ω impedance of the spectrum analyzer by a home-built low-noise high electron mobility transistor (HEMT) made using molecular beam epitaxy. These specially designed HEMT's can reach unprecedented low noise levels at 1 MHz with a noise voltage of 0.25nV/$\sqrt{Hz}$ and a noise current of 2.2fA/$\sqrt{Hz}$ and under deep cryogenic conditions (≤ 4.2 K), and with an input capacitance of about 5pF[23,24]. In addition, components $C_3$=10pF, $R_1$=10 Ω and $R_4$=10 Ω are placed close to the HEMT case to improve its stability.

The operation point of the HEMT is determined by $R_2$ and $R_3$ and the supply voltage. Since we aim to have a very low power dissipation we choose $R_2$=1kΩ to give a saturation current of the HEMT of a few tenths of mA. To ensure that the HEMT is in saturation we measured the drain current as function of drain-source voltage at room temperature and 4K by varying the supply voltage, as depicted in figure 3c. In the following demonstration the HEMT is biased in saturation at V=0.5V.

The voltage fluctuations in the 50 Ω line are amplified at room temperature by a +40dB current amplifier with an input current noise of 310 $pV/\sqrt{Hz}$ (Femto HAS-X-1-40) and is finally terminated by the 50 Ω input impedance of a Zurich Instruments MFLI digital spectrum analyzer. The power spectral density measured at the input of the spectrum analyzer is plotted in figure 3d where the blue dots are the measured data points and the red curve represents a circuit diagram fit.

**V. Noise spectroscopy performance on atomically Au(111)**

To demonstrate the simultaneous use of the STM feedback system and noise sensitive measurements in the tunneling regime, we performed noise spectroscopy measurements on a gold on mica sample. We believe that the Au(111) surface is most ideal for characterizing our noise-sensitive measurement since the sample is metallic thus any electron correlations are negligible. Figure 4a depicts an atomic-resolution image of the Au(111) terminated surface on a 20nm field of view, the characteristic 'herringbone' reconstruction is clearly visible.

In the same field of view we performed noise-spectroscopy measurements to resolve the current noise with atomic-scale resolution. Even though several topographic features can be observed, the spatially resolved noise map (figure 4b) exhibits homogeneous contrast, as is expected for a classical uncorrelated flow of electrons between sample and tip. The 64x64 pixel noise map was acquired in circa 12 hours.

The single point noise spectra (figure 4c) acquired at randomly chosen sites always show a linear increase of the noise with increasing current for a typical tunneling junction resistance (1GOhm). The linear increase of the noise is a unique characteristic of shot noise in the junction.

**VI. MHz differential conductance measurements**

A second application for our amplifier circuit is to measure differential conductance (*dI/dV*) which is proportional to the local density of states in a frequency range where 1/*f* noise is suppressed. In conventional STM differential conductance measurements, a voltage modulation has to be applied in the DC-1kHz range, now we can also perform dI/dV measurements at 3MHz. At this frequency, 1/f (and other) noise should be considerably lower, as we discussed in section II, figure 1a. Therefore, we expect that differential conductance measurements performed at 3MHz will show superior performance over conventional spectroscopic-imaging STM measurements.

Figure 5a shows that we can resolve the 3MHz voltage modulation, applied to the sample, within the tank bandwidth; the sharp peak in figure 5a (0.5 mV modulation amplitude at 3.009MHz) is 6 orders of magnitude higher than the background. To verify that the signal to noise ratio of MHz differential conductance measurements is higher than state-of-the-art STM techniques, we will compare the two by the means of a demonstration measurement on a Pb(111) sample using a Pb-coated PtIr tip at 3.3K.

First, we compare the differential conductance measured over a period of 10s in tunneling (6mV, 200pA) with the feedback disabled. In this way we can check the influence of slow (1/f) fluctuations on the differential conductance signal. Comparing the conventional dI/dV measurement (figure 5b) with the MHz measurement (figure 5c) reveals that the former is subject to much severe low-frequency fluctuations than the latter. This directly translates into a higher quality dI/dV spectrum, as can be seen by comparing the 3MHz dI/dV spectrum (figure 5e) to the one measured at 887Hz (figure 5d).

**Conclusion & outlook**

We have built a low temperature, low noise amplifier to measure STM currents at 3MHz. We used two superconducting Nb inductors to form a bias-tee and tank resonator coupled to a home-built, low-noise HEMT which is essential for the impedance matching to 50 Ω coax cable. We demonstrated the performance of this amplifier by performing noise spectroscopy measurements on an Au(111) surface, showcasing simultaneous visualization of the surface topology and atomically resolved noise maps. We further demonstrated that the amplifier allows to measure differential conductance spectra at 3MHz where 1/f noise is strongly suppressed.

We believe this newly developed technique will be useful for a variety of applications. Our immediate goal is to investigate many-body correlation effects in many quantum materials, including fluctuating stripes or orbital currents that have been proposed before[25,26], Kondo effects[27], or signature of Majorana modes[28,29]. Further, electron spin resonance often leads to periodic processes and equilibration times in the MHz to GHz regime[30,31]. These can be measured impedance matched with the presented amplifier instead of measuring indirect effects on the DC current. This could be further improved by guiding the microwave signal directly on the tip with coplanar waveguides, as suggested recently[32]. Finally, one could imagine that the thermal noise, introduced here as an unwanted noise source, could yield information about the sample via cross-correlation noise[33].

*Note: A very similar amplifier is reported by Massee et al.: "Atomic scale shot-noise using broadband scanning tunnelling microscopy".*

### Acknowledgements

We thank Ram Aluru, Marco Aprilli, Irene Battisti, Sander Blok, Kier Heeck, Maarten Leeuwenhoek, Freek Massee, Kees van Oosten, Tjerk Oosterkamp, Marcel Rost, Jan van Ruitenbeek and Gijsbert Verdoes for valuable discussions. This project was financially supported by the European Research Council (ERC StG SpinMelt) and by the Netherlands Organisation for Scientific Research (NWO/OCW), as part of the Frontiers of Nanoscience program, as well as through Vidi (680-47-536) and Projectruimte (16PR1028) grants.

16. Mamin, H. J., Birk, H., Wimmer, P. & Rugar, D. High-speed scanning tunneling microscopy: Principles and applications. *J. Appl. Phys.* **75,** 161 (1994).

17. Burtzlaff, A., Weismann, A., Brandbyge, M. & Berndt, R. Shot Noise as a Probe of Spin-Polarized Transport through Single Atoms. *Phys. Rev. Lett.* **114,** 016602 (2015).

18. Burtzlaff, A., Schneider, N. L., Weismann, A. & Berndt, R. Shot noise from single atom contacts in a scanning tunneling microscope. *Surf. Sci.* **643,** 10 (2016).

19. DiCarlo, L. *et al.* System for measuring auto- and cross correlation of current noise at low temperatures. *Rev. Sci. Instrum.* **77,** 073906 (2006).

20. Arakawa, T., Nishihara, Y., Maeda, M., Norimoto, S. & Kobayashi, K. Cryogenic amplifier for shot noise measurement at 20 mK. *Appl. Phys. Lett.* **103,** 172104 (2013).

21. Hashisaka, M. *et al.* Measurement for quantum shot noise in a quantum point contact at low temperatures. *J. Phys. Conf. Ser.* **109,** 012013 (2008).

22. Robinson, A.M. & Talyanskii, V.I. Cryogenic amplifier for 1 MHz with a high input impedance using a commercial pseudomorphic high electron mobility transistor. *Rev. Sci. Instrum.* **75,** 3169 (2004).

23. Dong, Q. *et al.* Ultra-low noise high electron mobility transistors for high-impedance and low-frequency deep cryogenic readout electronics. *Appl. Phys. Lett.* **105,** 013504 (2014).

24. Jin, Y. *et al.* Ultra-low Noise CryoHEMTs for Cryogenic High-Impedance Readout Electronics : Results and Applications. *ICSICT Conference proceedings* (2016).

25. Carlson, E. W., Dahmen, K. A., Fradkin, E. & Kivelson, S. A. Hysteresis and noise from electronic nematicity in high-temperature superconductors. *Phys. Rev. Lett.* **96,** 097003 (2006).

26. Kivelson, S. A. *et al.* How to detect fluctuating order in the high-temperature superconductors. *Rev. Mod. Phys.* **75,** 1201 (2003).

27. Figgins, J. & Morr, D. K. Differential Conductance and Quantum Interference in Kondo Systems. *Phys. Rev. Lett.* **104,** 187202 (2010).

28. Nadj-Perge, S. *et al.* Observation of Majorana fermions in ferromagnetic atomic chains on a superconductor. *Science* **346,** 602 (2014).

29. Golub, A. & Horovitz, B. Shot noise in a Majorana fermion chain. *Phys. Rev. B.* **83,** 153415 (2011).

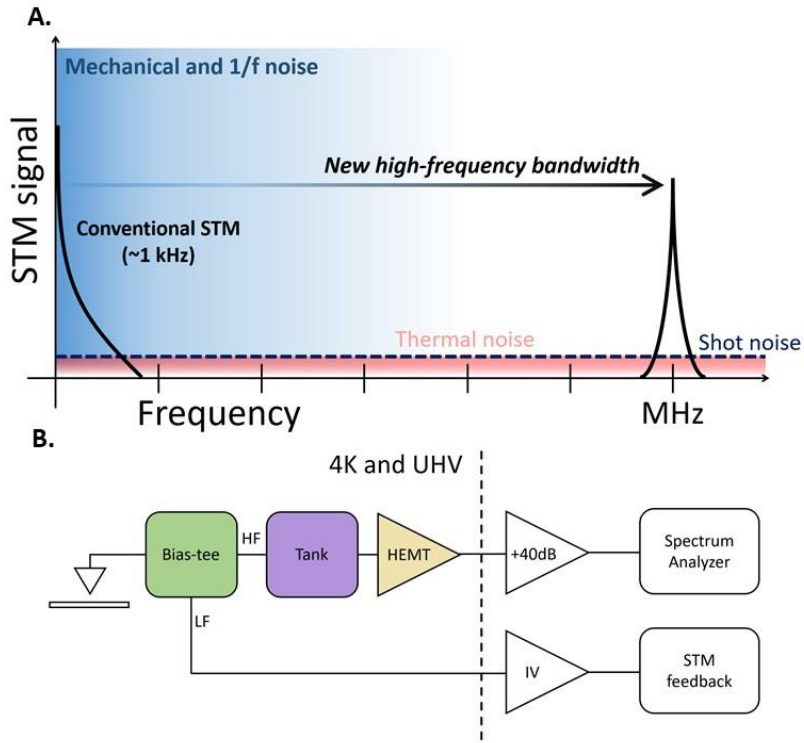

**Figure 1.** Noise in scanning tunneling microscopy (STM). **a**. The different noise sources in STM and their frequency dependence are depicted in this schematic plot. At low frequencies mechanical and 1/f noise dominate (indicated by blue region), in this region conventional STM is sensitive. To measure shot noise in the tunnel junction we need to create a new bandwidth at high-frequency. Here thermal noise and shot noise are the most dominant noise sources, since they are independent of frequency. **b.** Requirements for the newly built amplifier for combining STM and noise-spectroscopy. Crucial components are highlighted: i) the bias-tee (green) that separates the low and high frequency signals. ii) tank circuit (purple). iii) High electron mobility transistor (HEMT, indicated in yellow) to amplify the high-frequency signal. Both the low and high frequency signals have additional room temperature amplification and detection (white).

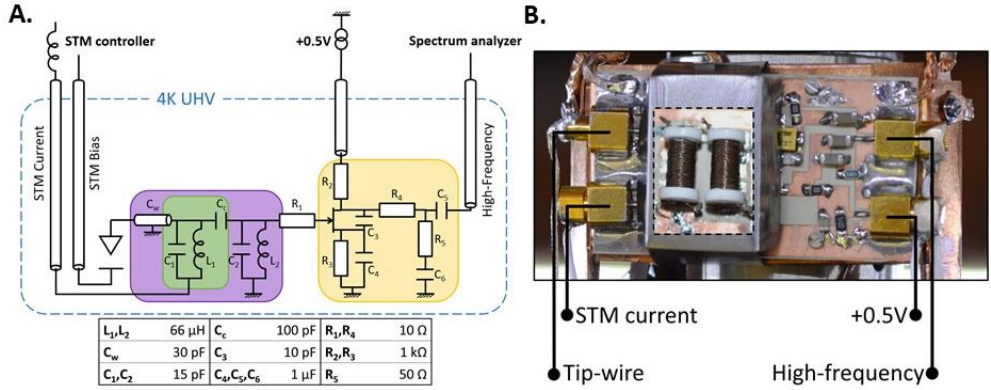

**Figure 2.** Newly developed amplifier for scanning noise spectroscopy. **a**. Circuit diagram of the amplifier. The colored boxes (green, purple and yellow) highlight specific parts of the amplifier corresponding to fig. 1b. **b.** Photograph of the printed circuit board. The niobium inductors are covered by a niobium shield.

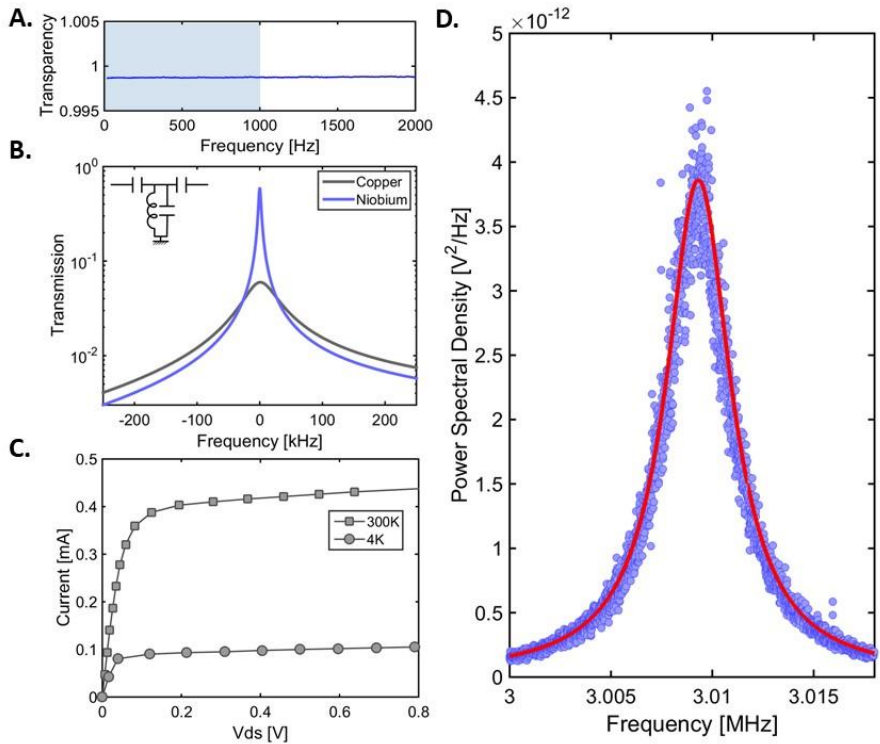

**Figure 3**. **a.** Low-frequency signal used for the STM feedback system. The transparency is close to 1 and flat from DC up to 2kHz. Bandwidth of the FEMTO transimpedance amplifier (1kHz) is indicated by the shaded blue area. **b.** Transmission of a home-built copper (grey) and superconducting niobium (blue) inductor resonator circuit. The latter showing a much higher quality factor. **c.** Current-voltage characteristics of the high electron mobility transistor (HEMT) at 300K and LHe temperatures. **d.** Power spectral density measured in a small bandwidth around the resonance frequency of the tank circuit (3.009 MHz). Blue dots are measured data points, red curve corresponds to a circuit diagram fit.

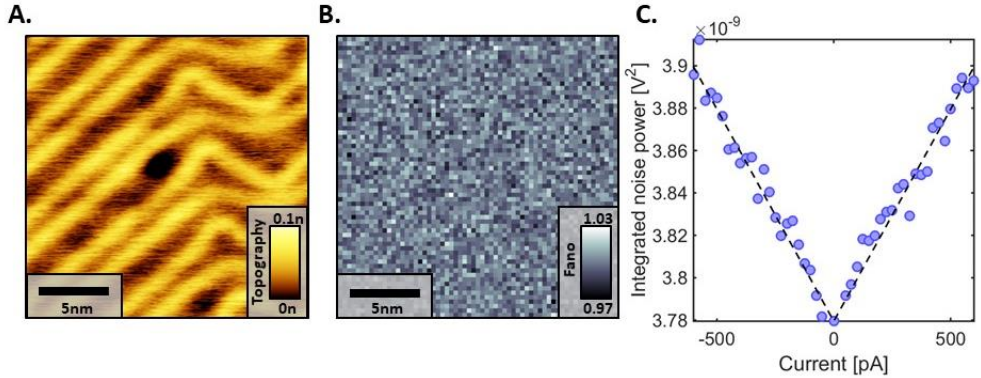

**Figure 4.** Benchmark of noise-sensitive measurements on a gold on mica sample. **a**. An atomically resolved STM topographic image of the Au(111) surface on a 20nm field of view, the 'herringbone' reconstruction is clearly visible. (Setup conditions bias: 100mV, current set point: 100pA). **b**. Spatially resolved noise map at 500mV, 500pA in the same field of view as Fig 4a, acquired in 12 hours. It shows homogenous Poissonian (Fano=1) noise at all locations. **c.** Single point noise spectrum acquired at a random location in Fig. 4b. The tunnel junction resistance is kept fixed to 1GOhm.

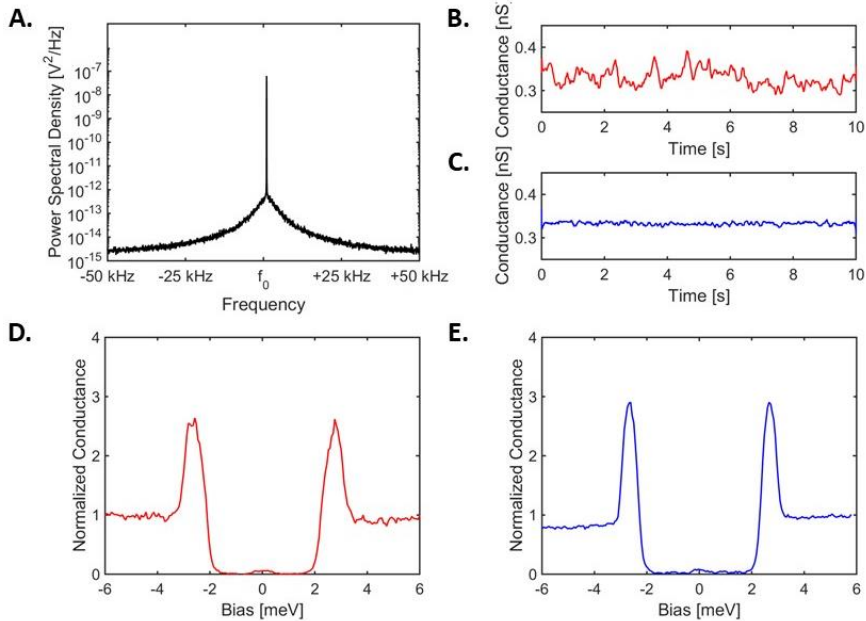

**Figure 5.** MHz differential conductance. **a.** Power spectrum obtained around the resonance peak. The modulation signal for the lock-in measurement is applied at the resonance frequency (3.009 MHz). **b.** Differential conductance measured over time at 887 Hz. A voltage modulation of 0.5mV is used and feedback is disabled. **c.** Differential conductance measured over time at the resonance frequency (3.009 MHz) under similar conditions as figure 5b. **d.** Differential conductance spectrum obtained with a 0.5mV voltage modulation at 887 Hz on a Pb(111) surface at a base temperature of 3.3K. **e.** Differential conductance spectrum obtained at the resonance frequency (3.009 MHz) obtained under the same conditions as figure 5d.

14